\documentclass[prb,aps,superscriptaddress,twocolumn,notitlepage,showpacs,nofootinbib]{revtex4-1}
\usepackage{latexsym}
\usepackage{amsmath, dsfont, physics}
\usepackage{amssymb}
\usepackage{graphicx}
\usepackage{caption}
\usepackage{subfigure}
\usepackage{float}
\usepackage{mathrsfs}
\usepackage{color}
\usepackage{txfonts}
\usepackage[justification=centering,
            format=plain]{caption}

\renewcommand{\raggedright}{\leftskip=0pt \rightskip=0pt plus 0cm}

\begin{document}

\title{Ultralow threshold bistability and generation of long-lived mode in a dissipatively coupled nonlinear system: application to magnonics}

\author{Jayakrishnan M. P. Nair}
\email{jayakrishnan00213@tamu.edu}
\affiliation{Institute for Quantum Science and Engineering, Texas A$\&$M University, College Station, TX 77843, USA}
\affiliation{Department of Physics and Astronomy, Texas A$\&$M University, College Station, TX 77843, USA}

\author{Debsuvra Mukhopadhyay}
\email{debsosu16@tamu.edu}
\affiliation{Institute for Quantum Science and Engineering, Texas A$\&$M University, College Station, TX 77843, USA}
\affiliation{Department of Physics and Astronomy, Texas A$\&$M University, College Station, TX 77843, USA}

\author{Girish S. Agarwal}
\email{girish.agarwal@tamu.edu}
\affiliation{Institute for Quantum Science and Engineering, Texas A$\&$M University, College Station, TX 77843, USA}
\affiliation{Department of Physics and Astronomy, Texas A$\&$M University, College Station, TX 77843, USA}
\affiliation{Department of Biological and Agricultural Engineering, Texas A$\&$M University, College Station, TX 77843, USA}
%\affiliation{Texas A$\&$M University, College Station, TX 77843, USA}

\date{\today}

\begin{abstract}
The prospect of a system possessing two or more stable states for a given excitation condition is of topical interest with applications in information processing networks. In this work, we establish the remote transfer of bistability from a nonlinear resource in a dissipatively coupled two-mode system. As a clear advantage over coherently coupled settings, the dissipative nature of interaction is found to support a lower pumping threshold for bistable signals. For comparable parameters, the bistability threshold for dissipatively coupled systems is lower by a factor of about five. The resulting hysteresis can be studied spectroscopically by applying a probe field through the waveguide and examining the polariton character of the transmitted field. Our model is generic, apropos of an extensive set of quantum systems, and we demonstrate our results in the context of magnonics where experimental interest has flourished of late. As a consequence of dissipative coupling and the nonlinearity, a long-lived mode emerges, which is responsible for heightened transmission levels and pronounced sensitivity in signal propagation through the fiber.  
\end{abstract}

\maketitle

\section{Introduction}
The manipulation of coherent coupling in hybrid quantum systems has been a cornerstone of research in quantum optics and information processing for several years. From classical to quantum, the unremitting quest for superior information processing algorithms and technologies has driven the field into the next dimension \cite{bennett2000quantum, wallquist2009hybrid, kimble2008quantum, RevModPhys.85.623, nielsen2002quantum}. Researchers were able to maneuver the dynamics of quantum systems in developing cutting-edge machinery for information processors, involving, for example, atoms \cite{bloch2012quantum}, trapped ions \cite{blatt2008entangled}, spins \cite{RevModPhys.79.1217}, superconducting circuits \cite{you2011atomic} and more. However, there exists no single all-encompassing quantum system capable of holding up to all the requirements and vital performance metrics of a modern-day signal-processing network. A photon, in spite of being a low-noise carrier of information, suffers from low storage potential. However, integrated photonic circuits \cite{xiang2013hybrid, treutlein2014hybrid}, which exploit strong light-matter interaction, seem to be progressively holding the leverage in the design of chip-scale information-processing devices with multitasking capabilities. 

Recently, hybrid optomagnonic systems utilizing ferrimagnetic materials like YIG have gained traction among the optics community \cite{PhysRevLett.111.127003, PhysRevLett.113.083603, PhysRevLett.113.156401, PhysRevLett.114.227201, zhang2015magnon, tabuchi2015coherent, chumak2015magnon, zhang2016cavity, PhysRevLett.116.223601, PhysRevLett.118.217201, zhang2017observation, PhysRevA.99.021801, PhysRevLett.121.203601, PhysRevResearch.1.023021, nair2020deterministic, PhysRevLett.120.057202, lachance2019hybrid, PhysRevB.102.104415}. YIGs are endowed with high spin density and the collective motion of these spins are embodied in the form of quasiparticles named magnons. Strong coherent coupling between photons and magnons has been used to realize an array of quantum and semiclassical effects including, but not restricted to, squeezing \cite{PhysRevA.99.021801}, entanglement \cite{PhysRevLett.121.203601, PhysRevResearch.1.023021, nair2020deterministic}, multistability \cite{PhysRevLett.120.057202, PhysRevB.102.104415}, exceptional points \cite{zhang2017observation} and dark modes \cite{zhang2015magnon}. In 2018, Harder \textit{et al} observed a dissipative form of magnon-photon coupling \cite{PhysRevLett.121.137203} underscored by the attractive nature of the eigenmodes, otherwise known as level crossing \cite{PhysRevB.99.134426, PhysRevApplied.11.054023, rao2019level, PhysRevB.100.094415, PhysRevLett.123.127202}.  While coherent coupling stems from the spatial overlap between two modes, dissipative coupling can be engineered by the inclusion of a shared reservoir (typically a waveguide) coupled independently to the two modes. Such an indirect coupling is mediated by a narrow bandwidth of propagating photons supported by the waveguide continuum, with dominant frequencies proximal to the mode transitions. Recently, Yu \textit{et al} \cite{PhysRevLett.123.227201} used an oscillator model to enunciate the physical origins of dissipative couplings. 

It is interesting to note that systematic studies on dissipatively coupled architectures with nonlinear components are far and few between. However, many systems that involve transmon qubits \cite{PhysRevA.76.042319} or magnonic excitations \cite{PhysRevB.94.224410} have intrinsic anharmonicity, which is often ignored as an inconsequential correction. The essence of our work is to drive home the imperative to examine the nonlinear domain of a dissipatively coupled two-mode system. The nonlinearity induces optical bistability, which finds applications as complex devices like photonic switches \cite{bilal2017bistable, li2014granular, babaee2016harnessing}, quantum memories \cite{PhysRevLett.113.074301, PhysRevLett.101.267203} etc. By taking up the model of a dissipative optomagnonic system interfacing with a waveguide channel, we lay bare the prospect of remote transfer of bistability from a spatially separated YIG sphere to a single-mode cavity. As a fruitful consequence, the dissipatively coupled system manifests remarkably low bistability threshold in contrast to a coherently coupled system with comparable coupling strength. The bistable signature is also mirrored by the modified polariton resonances of the waveguide transmission. In addition, the driven nonlinearity spawns a long-lived eigenmode with ultra-small linewidth leading to anomalous transmission effects like enhanced sensitivity in waveguide transmission, with potential applications in futuristic signal-processing networks.

This paper is organized as follows. In section \ref{sec1}, we develop a theoretical model to investigate the nonlinear response of a two-mode system coupled via an intermediary waveguide. We provide a detailed analysis of the bistability and the critical drive power required to achieve the same. In section \ref{sec2}, we explore the specific example of a driven cavity-magnon system. For experimentally realizable values of the system parameters, our simulations demonstrate bistability in the cavity response. In section \ref{sec3}, we employ a spectroscopic technique to advance a testable scheme for ratifying the bistability. Sec \ref{sec4} contains a concise description of the higher-dimensional eignensystem, which is directly responsible for a new domain of coherences.

\section{Theoretical Model}\label{sec1}
Before delving into any particular empirical models, we spell out the theoretical formalism describing a large class of two-mode anharmonic systems. The characteristic Hamiltonian goes as
\begin{equation}
\begin{split}
\mathcal{H}/\hbar= \omega_{a}a^{\dagger}a+\omega_{b}b^{\dagger}b+g(ab^{\dagger}+a^{\dagger}b)&\\+\mathcal{U}b^{\dagger 2}b^2+i\Omega (b^{\dagger}e^{-i \omega_d t}-b e^{i \omega_dt}),
\end{split}
\end{equation}
where $\omega_{a}$ and $\omega_{b}$ denote the respective resonance frequencies of the uncoupled modes $a$ and $b$, and $g$ signifies the direct coupling between them. The parameter $\mathcal{U}$ is a measure of the strength of Kerr nonlinearity intrinsic to the mode $b$ and conjured up by an external laser (pump) at frequency $\omega_d$, for which $\Omega$ signifies the Rabi frequency. While the Hamiltonian captures the effect of coherent coupling between the modes, a an embodiment of the dissipative coupling introduced by an interposing reservoir entails full recourse to the master-equation formalism. %In addition, these modes could be interfacing with a dissipative environment. Dissipative environments in an open quantum system fall roughly under two classifications - one, where the modes are independently coupled to their heat baths, and another, where a common reservoir interacts with both, as depicted in figure ().
For a two-mode system with a density matrix $\rho$, the corresponding the master equation would be given by
\begin{equation}
\frac{\dd \rho}{\dd t}=-\frac{i}{\hbar}[\mathcal{H},\rho]+\gamma_{a}\mathcal{L}(a)\rho+\gamma_{b}\mathcal{L}(b)\rho+2\Gamma \mathcal{L}(c)\rho,
\end{equation}
where $\gamma_{a}$ and $\gamma_{b}$ are the intrinsic damping rates of the modes respectively, while the parameter $\tau$ is tied to coherences introduced by the common reservoir. $\mathcal{L}$ is the Liouillian defined by its action $\mathcal{L}(\sigma)\rho=2\sigma\rho \sigma^{\dagger}-\sigma^{\dagger}\sigma\rho-\rho \sigma^{\dagger}\sigma$ and $c=\frac{1}{\sqrt{2}}(a+e^{i\phi}b)$ is the jump operator for symmetrical couplings to the reservoir. Here $\phi=2\pi L/\lambda_{0}$ symbolizes the phase difference between the two couplings, where $\lambda_{0}$ represents the resonant wavelength and $L$ denotes the spatial separation between $a$ and $b$. Assuming the wavelength of the resonant mode to be much bigger than the spatial separation, it makes sense to approximate $\phi\approx 0$. In the rotating frame of the laser drive, the mean value equations for $a$ and $b$ would be obtained as
\begin{figure}
 \captionsetup{justification=raggedright,singlelinecheck=false}
 \centering
   \includegraphics[scale=0.45]{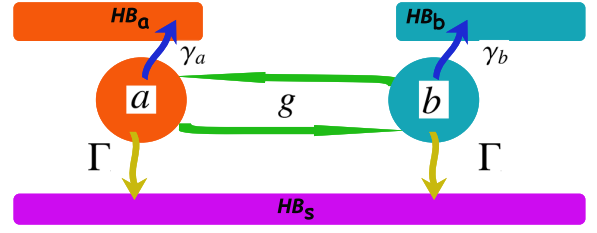}
\caption{General two-mode system, with both coherent and dissipative pathways of interaction. $HB_{\alpha}=$ heat bath of mode $\alpha$, with $\alpha=S$ denoting a shared reservoir.}
\label{sch}
\end{figure}
 \begin{equation}
 \begin{split}
 \begin{pmatrix}
\dot{a} \\
\dot{b}
\end{pmatrix} =-i\begin{pmatrix}
\delta_{a}-i(\gamma_{a}+\Gamma) & g-i\Gamma \\
g-i\Gamma & \delta_{b}-i(\gamma_{b}+\Gamma) 
\end{pmatrix}\begin{pmatrix}
a \\
b
\end{pmatrix} &\\+2\mathcal{U}(b^{\dagger}b)\begin{pmatrix}
0 & 0 \\
0 & 1 
\end{pmatrix}\begin{pmatrix}
a \\
b
\end{pmatrix}+\begin{pmatrix}
0\\
\Omega
\end{pmatrix},
\end{split}
 \end{equation}
where $\delta_i=\omega_i-\omega_d$ ($i=a,b$). For brevity, $\expval{...}$ notations have been excluded. Invoking the mean-field approximation allows one to decouple the higher-order expectations as $\expval{\xi_1\xi_2}=\expval{\xi_1}\expval{\xi_2}$, so that $\expval{b^{\dagger}bb}$ essentially reduces to $\abs{b}^2b$.

Purely dissipative models can be pinned down by the choices $g=0$ and $\Gamma\neq 0$. Within this category, one can achieve an anti-PT symmetric mode hybridization by tuning the control field to a frequency such that $\delta_b=-\delta_a=\delta/2$ and by enforcing equal damping rates $\gamma_a=\gamma_b=\gamma_0$. Such models then satisfy $\{\hat{PT}, \mathcal{H}\}=0$, where $\{.,.\}$ stands for the anti-commutator. When the stability criterion is fulfilled, the time-dependent solutions decay into a stationary value in the long-time limit, i.e. $(a, b) \rightarrow (a_0, b_0)$, which can be easily derived from Eq. (3) by setting $\dot{a}=\dot{b}=0$. These stationary values are, therefore, entwined via the constraints
\begin{align}
(2\gamma-i\delta)a_0+2\Gamma b_0=0,\notag\\
(2\gamma+i\delta)b_0+4i\mathcal{U}\abs{b_0}^2b_0+ 2\Gamma a_0-2\Omega=0,
\end{align}
where $\gamma=\gamma_0+\Gamma$. So $\gamma_0$ represents the spontaneous rate of emission into the local heat baths. Eliminating $b_0$, one has a polynomial equation for $y=\abs{a_0}^2$, predicting bistability in the ensuing response,
\begin{align}
\frac{\beta^2}{\Gamma^{2}}y-2\mathcal{U}\beta\delta\frac{ \gamma^2+(\delta/2)^2}{\Gamma^4}y^2+4\mathcal{U}^2\frac{(\gamma^2+(\delta/2)^2)^{3}}{\Gamma^{6}}y^3=I,
\end{align}
with the definitions $\beta=\Gamma^2-\gamma^2-(\delta/2)^2$ and $I=\Omega^2$. Note, \textit{en passant}, that this bistability is not ingrained in mode $a$, rather, it is transferred from the anharmonic mode $b$. The turning points of the bistability curve can be gleaned from the expression above by inspecting the solutions to $\frac{\dd I}{\dd y}=0$, which turn out to be
\begin{align}
y_{\pm}=\frac{\beta \Gamma^{2}}{6\mathcal{U}[\gamma^2+(\delta/2)^2]^{2}}\bigg[{\delta}\pm\sqrt{(\delta/2)^2-3\gamma^2}\bigg].
\end{align}
Thus, the condition for observing bistable signature can be encoded as: $\mathcal{U}\delta<0$ and $\delta^2>12\gamma^2$. In addition, there is a cutoff value for the pump power beyond which the bistable characteristics set in. The appropriate magnitude of $I^{(c)}$ pertains to the inflection point in the $I-y$ graph and can be inferred from the conditions $\frac{\dd I}{\dd y}=\frac{\dd^2 I}{\dd y^2}=0$. For any generic two-mode system with the interaction comprising both coherent and dissipative components, i.e. $g\neq 0, \Gamma\neq 0$, the cutoff power obeys the relation
\begin{align}
I^{(c)}= \frac{1}{432}\frac{[\sqrt{3}(4\gamma^{2}-\Gamma^{2}+g^{2})\text{sgn}(\mathcal{U})+2g\Gamma]^{3}}{\mathcal{U}\gamma^3},
\end{align} 
where sgn(x) is $+1$ when $x>0$ and $-1$ for $x<0$. For all the subsequent analyses in this paper, we shall work with the assumption $\mathcal{U}>0$. The precise ramifications of the expression (7) in relation to the nature of coupling can be understood by comparing the cutoffs for coherently coupled and dissipatively coupled systems by letting $\Gamma=0$ and $g=0$ respectively. This leaves us with 
\begin{align}
\frac{I_{dis}^{(c)}}{I_{coh}^{(c)}}=\bigg(\frac{4\gamma^{2}-\Gamma^{2}}{4\gamma^{2}+g^{2}}\bigg)^{3},
\end{align}
where we have assumed that the two kinds of systems decay effectively at the same rate. Note that for $\Gamma\neq 0$ and $g\neq 0$, the expression above is always less than one. Stated differently, a dissipatively coupled system demonstrates a lower bistability threshold when compared to a coherently coupled system with identical coupling strength. More importantly, we can achieve an extremely low threshold in a dissipatively coupled system by signifiicantly boosting the waveguide-mediated interaction relative to the intrinsic damping, such that $\gamma \approx \Gamma$. To put this into perspective, for two systems with identical coupling strengths, that is $\Gamma \approx g$, the quantity in Eq. (8) goes over to a finite value equaling $0.22$ as $\gamma\rightarrow \Gamma$. Despite the fact that a dissipatively coupled system showcases remarkably low bistability threshold, the $78\%$ advantage is the best that can be attained in an experimental apparatus. 

\section{Dissipatively coupled nonlinear magnonic system}\label{sec2}
\begin{figure}
 \captionsetup{justification=raggedright,singlelinecheck=false}
 \centering
   \includegraphics[scale=0.45]{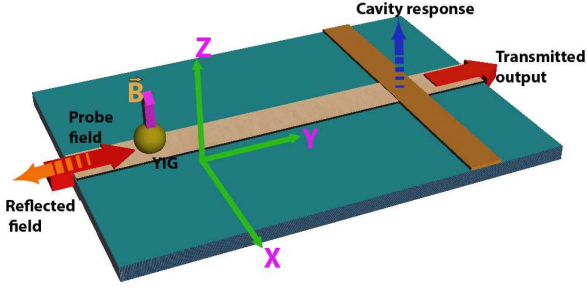}
\caption{Schematic of a YIG sphere interacting remotely with a single-mode cavity, with the coupling mediated by a waveguide.}
\label{sch}
\end{figure}
To give substance to the foregoing discussion on dissipatively coupled systems, we consider a hybrid optomagnonic apparatus that has already evolved into a topical system of experimental activities. Our setup consists of a single-mode optical cavity and a YIG sphere separated by some finite distance and interacting dissipatively with each other via an intermediary one-dimensional waveguide, as illustrated in fig. (2). The intrinsic anharmonicity of the YIG \cite{PhysRevB.94.224410} is kindled by a microwave laser acting as the pump. The Hamiltonian of this driven system can be laid out as
\begin{equation}
\begin{split}
\mathcal{H}/\hbar = -\gamma_e & B_{0} S_{z} + \gamma_e^2\frac{\hbar K_{\text{an}}}{M^2 V}S_{z}^2 + \omega_a a^{\dagger}a +\mathcal{H}_{\Omega},
\end{split}
\label{h}
\end{equation}
 where $\mathcal{H}_{\Omega}=i\hbar\Omega\left(m^{\dagger} e^{-i\omega_{\text{d}}t} - m e^{i\omega_{\text{d}}t}\right)$, $B_{0}$ is the static magnetic field applied along the $+\hat{z}$ direction, $\gamma_e=e/{m_e c}$ is the gyromagnetic ratio for electron spin and $\textbf{S}$ the collective spin operator of the YIG.  The quantity $\textbf{M}=\gamma_e\textbf{S}/V$ denotes the magnetization, where $V$ is the volume of the YIG. The second term in Eq. (9) stems from the magnetocrystalline anisotropy. The parameter $\omega_{a}$ represents the cavity resonance frequency and $a$ $(a^{\dagger})$ characterizes the annihilation (creation) operator of the cavity. Invoking the Holstein-Primakoff transformation \cite{PhysRev.58.1098} in the limit of large spin, the above Hamiltonian simplifies to Eq. (1) with $g=0$, frequency $\omega_{b}= \gamma_e B_{0}-\frac{2\hbar\gamma_e^2 S K_{\text{an}}}{M^2 V}$, $\mathcal{U}=\frac{\gamma_e^2K_{\text{an}}}{M^2 V}$. The Bosonic operators $(b, b^{\dagger})$, which represent the magnonic quasiparticle, are related to the spin operators as $S^+ =\sqrt{2S}b, S^- =\sqrt{2S}b^{\dagger}$ in the limit $S\gg\langle b^{\dagger}b\rangle$. Here, $S=\frac{5}{2}\rho V$ is the collective spin of the YIG, with the Fe$^{3+}$ ion density and the diameter of the YIG given by  $\rho=4.22\times 10^{27}$ m$^{-3}$ and ${d}=1$ mm respectively.
\begin{figure}
 \captionsetup{justification=raggedright,singlelinecheck=false}
 \centering
   \includegraphics[scale=0.5]{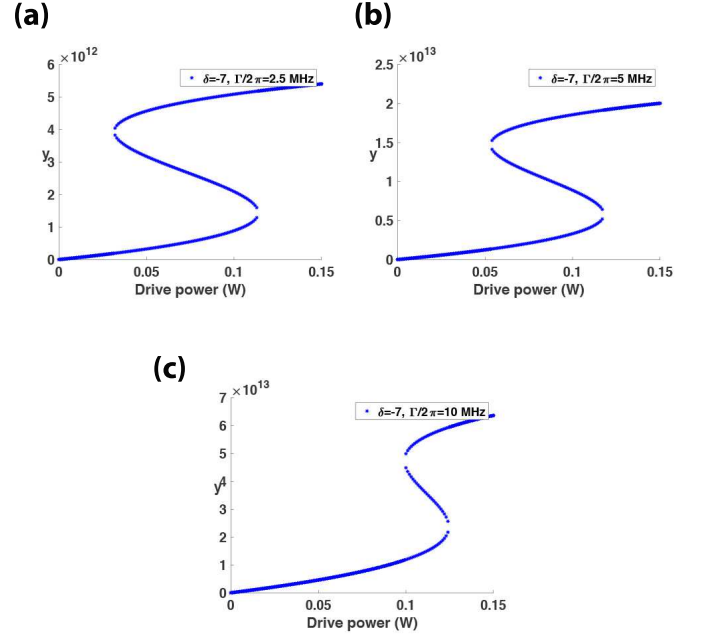}
\caption{Response in the cavity field plotted against the pump power, for varying strengths of the dissipative coupling. Weaker dissipative couplings support larger ranges of the drive power pertaining to bistable states in the response. The parameters $\gamma_{0}/2\pi=5$ MHz and $\mathcal{U}/2\pi=42.1$ nHz.}
\label{sch}
\end{figure}
The Rabi frequency $\Omega=\gamma_e\sqrt{\frac{5\pi\rho d D_{\text{p}}}{3c}}$ is determined by the system characteristics as well as the pump power $D_p$. Here, $c$ stands for the speed of light. As discussed in the general setting, the cavity signal satisfies the multistable equation expressed in Eq. (5). In figure 3(a), we plot the optical response against various pump powers. As we slowly ratchet up the drive power, an abrupt jump is observed in the signal as opposed to the fairly linear ascent in the domain of weaker drives. A similar precipitous transition is seen as we tamp down the drive power, this time at a different point, unraveling the transfer of bistability to the cavity. This is analogous to the optical bistability reported in a coherently coupled cavity-magnon system \cite{PhysRevLett.120.057202}, in which the nonlinear resource (YIG) is contained in the cavity. However, the paradigm we explore has the antithetical underpinnings of spatial separation being the source of coupling, as the YIG is only remotely interacting with the cavity. A boost in the dissipative coupling brings about a diminishing window of bistability, as illustrated in figure 3(b)-(c). This bears stark disparities with coherently coupled systems, where a stronger coupling widens the region of bistability. Nevertheless, this feature turns out to be quite advantageous as we can generate broader bistability windows for modest values of the engineered dissipative coupling. The observed bistability is fairly robust against intrinsic damping rates of the modes, as evident from the figures 4(a)-(d). 

It makes for a relevant observation that as the dissipative coupling is downsized, not only is the bistability window broadened but the effect is also rendered accessible at lower drive powers. This seems to be a subtle point of divergence between directly and indirectly coupled systems. That said, one cannot, of course, make the dissipative coupling indefinitely weak and still expect to harness bistable signatures to one's advantage. This is a trivial consequence of the fact that weak couplings cannot elicit strong responses from the cavity.

 \begin{figure}
 \captionsetup{justification=raggedright,singlelinecheck=false}
 \centering
   \includegraphics[scale=0.5]{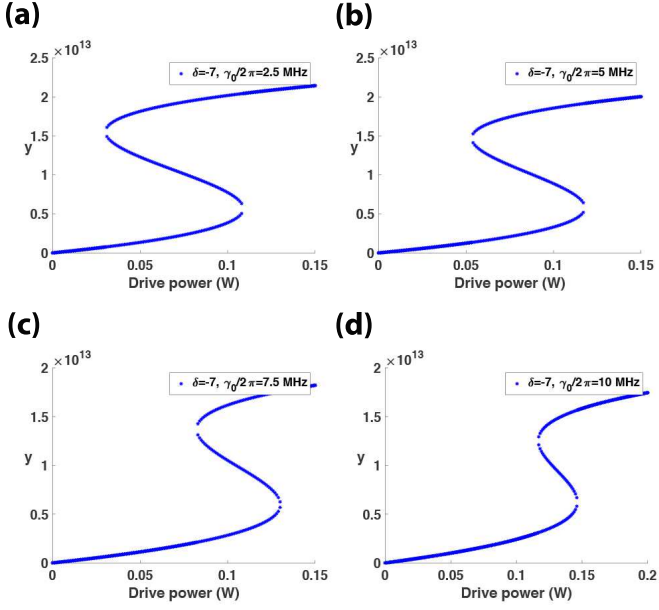}
\caption{Effect of intrinsic damping on the optical bistability. Other parameters are $\Gamma/2\pi=5$ MHz and $\mathcal{U}/2\pi=42.1$ nHz. The detunings are scaled down by a factor $S=2\pi\cross 20$ MHz. }
\label{sch}
\end{figure}
 \section{Observation of bistability via the waveguide transmission}\label{sec3}
 \begin{figure}
 \captionsetup{justification=raggedright,singlelinecheck=false}
 \centering
   \includegraphics[scale=0.5]{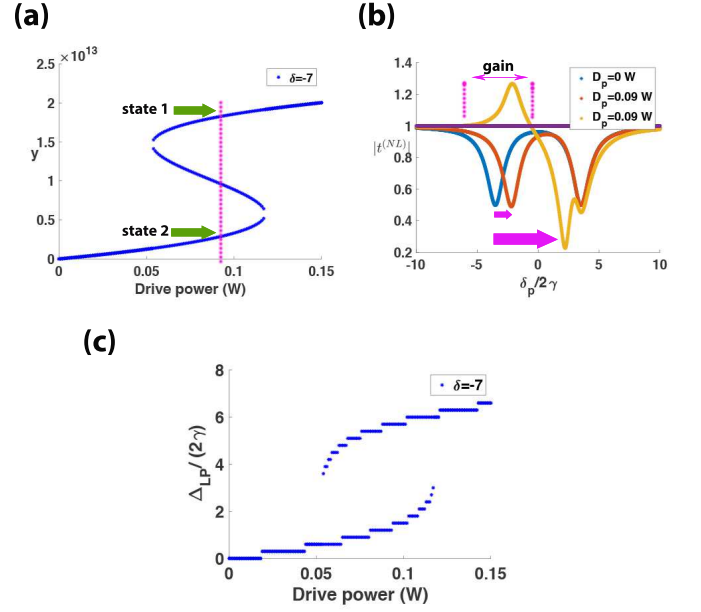}
\caption{(a) The cavity response plotted against the pump power showcasing the two stable states at $D_p=90$ mW. (b) Transmission spectra through the waveguide as a function of the scanning frequency. (c) Shifts in lower polariton frequency vs. the pump power $D_p$. Other parameters are $\Gamma/2\pi=5$ MHz, $\gamma_{0}/2\pi=5$ MHz and $\mathcal{U}/2\pi=42.1$ nHz. Scaling factor for detunings $S=2\pi\cross 20$ MHz.}
\label{sch}
\end{figure}
 Spectroscopy is a quintessential tool in science and engineering, and is routinely applied to QED systems. The fundamental spectroscopic principle of is to probe the system using a weak electromagnetic excitation, and from the transmission properties of the system, one extracts key information about the system, including, but not limited to, its eigenmode configuration. Here, we employ a similar technique to investigate the transmission properties of the nonlinear optomagnonic device with varying drive powers, as depicted in figure (2). Specifically, we show how anharmonicity-induced shifts in the polariton minima reinforce the bistable nature of cavity signal. In the presence of a monochromatic probe field guided through the fiber, the Hamiltonian in Eq. (1) gets modified to $\mathcal{H}+\varepsilon\mathcal{H'}$, where $\mathcal{H'}=i\hbar[(a^{\dagger}+b^{\dagger}) e^{-i\delta_{\text{p}}t} - h.c.)]$, $\delta_{p}=\omega_{p}-\omega_{d}$, $\varepsilon=\sqrt{2\Gamma \mathcal{P}_{\varepsilon}/\hbar \omega_{p}}$ and $\mathcal{P}_{\varepsilon}$ is the probe power. The updated Langevin equations then entail 
 \begin{align}
& \dot{a} = -(-i\delta/2 + \gamma)a -2\Gamma b + \varepsilon e^{-i\delta_{p}t},\notag \\
& \dot{b} = -(i\delta/2 + \gamma)b - 2i \mathcal{U} \abs{b}^2b -\Gamma a + \Omega + \varepsilon e^{-i\delta_{p}t}.
\label{QLE}
\end{align}
The solution to this set of equations in the long-time limit can be written as a Fourier series expansion,
\begin{equation}
a= \sum_{n=-\infty}^{\infty} a_{(n)} e^{-in\delta_{p} t},\quad b = \sum_{n=-\infty}^{\infty} b_{(n)} e^{-in\delta_{p} t},
\label{FE}
\end{equation}
where $a_{(n)}$ and $b_{(n)}$ are the amplitudes associated with $n$-th harmonic of the probe frequency. The steady components $a_0, b_0$, which are actually oscillating at the pump frequency, conform to Eq. (4). The probe being a weak field, we ignore the higher-order terms and truncate the series with $n$ running up to $\pm 1$. Utilizing Eqs. (13) and (14), we conclude the following set of linear equations for the oscillating components of the steady state:
\begin{align}
\mathcal{M}X^{(1)}=F_{\varepsilon},
\end{align}where $\mathcal{M}=
  \begin{pmatrix}
\mathcal{A} & \mathcal{B}\\
\mathcal{B^*} & \mathcal{C}
  \end{pmatrix}$,  $X^{(1)}=( a_{+}, b_{+}, a_{-}^{*}, b_{-}^{*})^T$, $F_{\varepsilon}=\varepsilon(1, 1, 0, 0)^T$, with the elements $\mathcal{A}=\begin{pmatrix}
\gamma-i(\delta/2+\delta_p) & \Gamma\\
\Gamma & -\gamma+i(\tilde{\Delta}+\delta_p)\\
\end{pmatrix}$, $\mathcal{B}=\begin{pmatrix}
0 & 0\\
0 & 2i\mathcal{U}b_0^2
\end{pmatrix}$,  $\mathcal{C}=\begin{pmatrix}
\gamma-i(\delta/2-\delta_p)) & \Gamma\\
\Gamma & \gamma - i(\tilde{\Delta}+\delta_p) \\
\end{pmatrix}$, and $\tilde{\Delta}=\frac{\delta}{2}+4\mathcal{U}\abs{b_{0}}^2$. The first-order fluctuations about the steady state can then be obtained numerically by inverting Eq. (12). Using the input-output relation $\varepsilon_t=\varepsilon-2\Gamma (a+m)$, where $\varepsilon_t$ designates the transmitted signal, we obtain the transmission coefficient at the probe frequency, 
  \begin{align}
t^{(NL)}=\varepsilon_{t}/\varepsilon &=1 - 2\Gamma(a_{+}+m_{+})/\varepsilon\notag\\
&=1+2\Gamma\sum_{r,s=1}^2(\mathcal{M}^{-1})_{rs}.
\end{align}
\begin{figure}
 \captionsetup{justification=raggedright,singlelinecheck=false}
 \centering
   \includegraphics[scale=0.5]{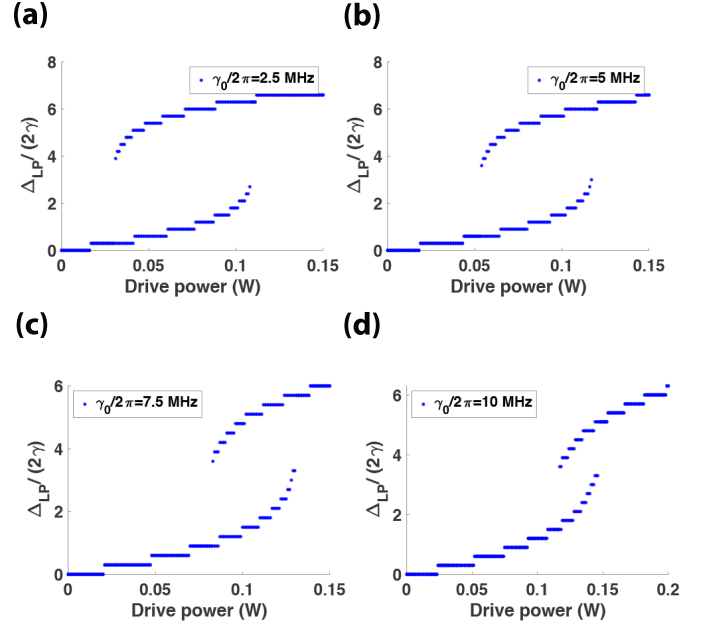}
\caption{The lower polariton frequency shifts against pump power $D_p$ for varying values of the intrinsic damping. All the other parameters are identical to figure (4).}
\label{sch}
\end{figure}
In the absence of a pump, when nonlinear effects remain dormant, the Rabi splitting between the cavity-magnon polariton branches is substantiated in fig. 5 (a). For a nonzero drive power, under conditions of bistability, the higher polariton resonance remains largely unscathed while the lower polariton sustains appreciable frequency shifts. Here, lower and higher polaritons (LP and HP) refer to the polariton branches with lower (appearing on the left) and higher (appearing on the right) frequencies in the transmission spectra. The respective minima are labelled as $\delta_{\text{LP}}$ and $\delta_{\text{HP}}$. The bistability in transmission is illustrated with two stable transmission spectra in figure 5(a) at drive power 0.09 W. In addition, we draw information on the hysteresis curve in figure 3 by investigating shifts in the lower polariton minima. We numerically obtain the corresponding shifts in polariton frequency $\Delta_{\text{LP}}=\delta_{\text{LP}}$-$\delta_{\text{LP}}^{(0)}$ as a function of input (pump) power, where $\delta_{\text{LP}}^{(0)}$ is the detuning of the lower branch at zero drive power. The bistability gets manifested in the frequency shifts in figure 5 (c), which replicate the pattern of the spincurrent curve in figure 3 (b). Further, we provide the frequency-shift curves (figures 6 (a)-(d)) against the intrinsic decay parameters of the cavity and magnon modes, testifying to the robustness of bistability against extraneous decoherence.

\section{Anharmonicity-induced long-lived mode} \label{sec4}
In this section, we show how the presence of nonlinearity activates new coherences, which, in turn, introduces optical anomalies in the fiber transmission. Owing to its nonlinear nature, the system of coupled-mode equations in (13) does not yield readily to a Hamiltonian-based analysis. However, an effective Hamiltonian can be eked out by appealing to a linearized approximation about the steady state, \textit{viz.} $a(t)=a_0+\delta a(t)$, $b(t)=b_0+\delta b(t)$. The fluctuations $\delta a(t)$ and $\delta b(t)$ are presumed to be general, albeit small in relation to $a_0$ and $b_0$. This permits the dismissal of higher-order effects in these variations, akin to what was invoked in the last section. The variables $\delta b(t)$ and $\delta b^{\dagger}(t)$ get interlinked due to the anharmonicity. The inter-coupling is, of course, too weak to bear on any observable effects at smaller drive powers. However, a drive power around $0.02$ $W$ makes this coupling paramount. Defining $Y(t)=Y+\delta\xi(t)$, where $Y(t)={\begin{pmatrix}
a(t) & b(t) & a^{\dagger}(t) & b^{\dagger}(t)\\
\end{pmatrix}}^T$ and $Y={\begin{pmatrix}
a & b & a^{\dagger} & b^{\dagger}\\
\end{pmatrix}}^T$ are both 4-element vectors, the Langevin equations reduce to a linear dynamical model, $\bigg[\dfrac{\dd}{\dd t}+i\mathcal{H}_{NL}\bigg]\delta\xi(t)=\mathcal{E}_{in}(t)$, with
\begin{align}
\mathcal{H}_{NL}=\begin{pmatrix}
-\frac{\delta}{2}-i\gamma & -i\Gamma &0 &0\\
-i\Gamma  &\tilde{\Delta}-i\gamma & 0 & 2\mathcal{U}b_{0}^2\\
0& 0& \frac{\delta}{2}-i\gamma & -i\Gamma \\
0 & -2\mathcal{U}b_{0}^{*2} & -i\Gamma &  -\tilde{\Delta}-2i\gamma
\end{pmatrix},
\end{align}
$\mathcal{E}_{in}(t)=\varepsilon{\begin{pmatrix}
e^{-i\delta_pt} & e^{-i\delta_pt} & e^{i\delta_pt} & e^{i\delta_pt}\\
\end{pmatrix}}^{T}$ 
\begin{figure}
 \captionsetup{justification=raggedright,singlelinecheck=false}
 \centering
   \includegraphics[scale=0.5]{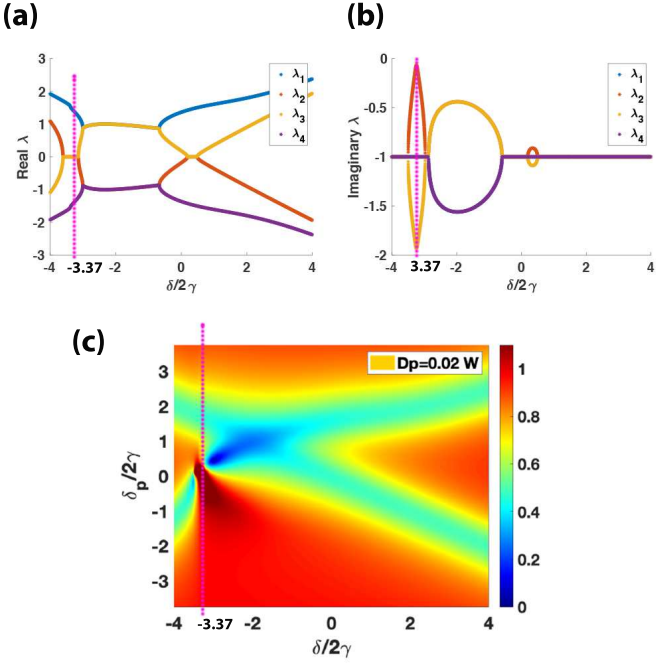}
\caption{(a) Real parts and (b) Imaginary parts of the eigenvalues of $\mathcal{H}_{NL}$ plotted against $\delta$ at $Dp=0.02$ W. (c) $|t|$ is plotted against $\delta_{p}$ and $\delta$ at the same pump power. The vertical lines in (a), (b) and (c), all at $\delta/2\gamma=-3.37$, identify the exceptionally long-lived mode with extreme linewidth suppression. Other parameters are identical to figure (5).}
\label{sch}
\end{figure}
and $\tilde{\Delta}$ has been defined in relation to Eq. (12) earlier. A straightforward comparison with Eq. (12) reveals that $\mathcal{M}=i(\mathcal{H}_{NL}-\delta_p)$, implying that anharmonic resonances of the system are tied to the eigenmodes of $\mathcal{H}_{NL}$. Any exotic properties of eigenmodes would, therefore, be translated to observable anomalies in the transmission signal. The eigenfrequencies denoted as $\lambda_i$'s, with $i$ running from $1$ through $4$, manifest characteristics attributable to level attraction. This is portrayed, for $D_p=0.02$ W and $\mathcal{U}/2\pi=42.1$ nHz in fig. 7(a), (b). Since the effective dimensionality of the Hilbert space is augmented by the introduction of anharmonicity, a major impact of $\mathcal{U}$ is to bring in new coherent phenomena in the transmission across the hybridized cavity-magnon system. As figures 7(c) and 8(a)-(d) reveal, the nonlinear coupling categorically skews the transmission lineshapes by introducing asymmetry. However, the case of negative magnon detunings turns out to be particularly intriguing, as a new transparency window crops up, which owes its origin to a long-lived mode of the system. 
\begin{figure}
 \captionsetup{justification=raggedright,singlelinecheck=false}
 \centering
   \includegraphics[scale=0.45]{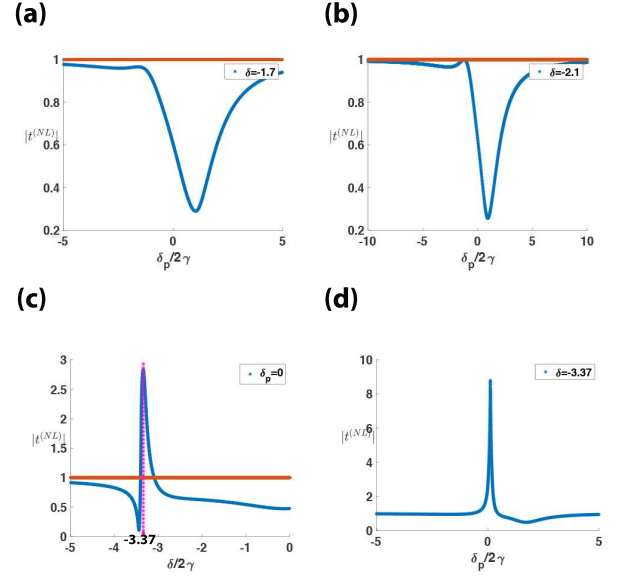}
\caption{Figures (a), (b) represent the transmission spectra against the probe detuning $\delta_{p}$ at two different values of $\delta$, portraying the asymmetry and the transition from unidirectional reflection to perfect transmission. (c) Transmission spectra against $\delta$ with $\delta_{p}=0$, highlighting a region of significant gain around $\delta/2\gamma=-3.7$. (d) Transmission plot demonstrating pronounced sensitivity around $\delta_{p}=0$. Other parameters are identical to figure (5). Scaling factor for detunings $S=2\pi\cross 20$ MHz.}
\label{sch} 
\end{figure}
With a significant Rabi frequency ($D_p\sim$ $0.02$ W), the enhanced pumping rate is brought to bear on the linewidths and one of them shrinks close to nought, leading to a region of highly elevated transmission levels. This is highlighted by a dark red smeared out streak shown in fig. 7(c), and also illustrated for better clarity in figs. 8(a), (b) where the transmission is plotted for a pair of fixed values of delta. The near-perfect transmission comes to light most prominently around $\delta/2\gamma=-2.1$, for the chosen set of system parameters. In fact, at this precise magnitude of the detuning, the waveguide photon gets transmitted unimpeded, demonstrating perfect transparency. Equally interesting is the collateral emergence of a strongly reflecting polariton minimum. Such fundamentally conflicting behaviors transpire in a narrow neighborhood around $\delta_p=0$ across which the device flips between a state of unidirectional reflection $\bigg(\abs{t^{(NL)}}\ll 1\bigg)$ to complete transparency $\bigg(\abs{t^{(NL)}}\approx 1\bigg)$. This conspicuous result suggests practical advantages in engineering driven systems for implementing nonlinearity-assisted strong signal switching.

 However, as evidenced by fig. 7(c), although the switching effect seems to be further amplified as $\delta$ becomes more strongly negative, we find an obtrusive region of pump-induced gain that makes the transmission surge past unity. Figure 8(c) depicts an extraordinarily high transmission peak around $\delta/2\gamma=-3.37$, at which the imaginary part of one of the poles (marked by yellow in 7(a), (b)) moves tantalizingly close to zero while the corresponding eigenfrequency remains approximately zero. This is why the signal dramatically shoots up around $\delta_p=0$. Additionally, it demonstrates a sharp sensitivity to the probe frequency, which is also responsible for the signal flipping effect. The spectacular sensitivity in the signal can be accredited to the existence of a second-order pole at $\delta_p\approx 0$, $\delta/2\gamma\approx -3.37$ in the derivative of the transmitted signal, i.e.
\begin{align}
\frac{\partial {t^{(NL)}}}{\partial \delta_p}=2i\Gamma\sum_{r,s=1}^2\{\mathcal{M}^{-2}\}_{rs}.
\end{align}
The above relation follows from the consideration that $\mathcal{M}^{-1}(\delta_p+\eta)-\mathcal{M}^{-1}(\delta_p)=i\eta\mathcal{M}^{-1}(\delta_p+\eta)\mathcal{M}^{-1}(\delta_p)$ and subsequently taking the limit $\eta\rightarrow 0$. One should, in practice, exercise caution by avoiding excessive gain, as any unduly long-lived mode with negligible width augurs a breakdown in the validity of linearized approximations to the steady state and requires a nonlinear treatment of the probe. Nonetheless, even if one avoids this unstable terrain of extraordinary transmission, one could carefully configure the detuning $\delta$ to enable transparent behavior and strong signal sensitivity near to $\delta_p=0$ as discernible from fig. 8(d). In this connection, we also note that sensitivity in dissipatively coupled systems has been studied in the context of linear perturbations \cite{zhang2020breaking}, and more recently, for the detection of weak anharmonicities in an optomagnonic model \cite{nair2020enhanced}.
%The transition from  to  begets a phenomenal flip from strong absorption to heightened supertransmission across the driven cavity-magnon system.
\section{Summary and Concluding Remarks}
To summarize, our foray into the nonlinear domain of dissipatively coupled models sheds light on two fundamental anharmonic signatures of driven systems: optical bistability and anharmonicity-induced coherences. Directly coupled nonlinear systems, such as a YIG sphere in a cavity, are already known to generate bistability. Our formulation demonstrates that dissipatively coupled models, like the one we study here, accords a lower pumping threshold for observing bistable states. In the context of an optomagnonic system, the frequency shifts in the polariton minima bring out the essential hallmarks of bistability, in lockstep with the cavity response. With the anharmonicity factored in, the increased dimensionality of an intricate but predominantly stable eigensystem imparts higher-order coherences to the hybridized polaritons, the significance of which is accentuated by an enhanced pumping rate. Simulation of the transmission across the waveguide reveals a parameter regime of pump-induced gain, where external energy is funneled into the fiber causing a coherent buildup of the output signal. The anharmonic gain is hardly an adventitious effect, but, rather, an upshot of intense linewidth inhibition in one of the resonances. This long-lived mode is also responsible for extreme sensitivity in the signal to the probe frequency. Fortuitously, the existence of a transmission window, flanking a polariton minimum, extends well into the stable regime. In this regime, the system demonstrates a stark duality in signal transport, marked by the possibility of achieving both transparency and opacity, for the same set of system parameters. The greater signal sensitivity and control incorporated by the anharmonicity could be harnessed for the design of photonic devices with optical switching properties.

\section{Acknowledgements}
The authors acknowledge the support of The Air Force Office of Scientific Research [AFOSR award no FA9550-20-1-0366], The Robert A. Welch Foundation [grant no A-1243] and the Herman F. Heep and Minnie Belle Heep Texas A\&M University endowed fund.

D.M. and J.M.P.N. contributed equally to this work.
\bibliography{references}

\end{document}